\documentclass[12pt]{article}
\usepackage{graphicx}
\usepackage{ulem}
\usepackage{geometry}
\usepackage{mathtools}
\usepackage{xcolor}
\usepackage{cancel}
\usepackage{amssymb}
\usepackage{mathrsfs}
\usepackage{soul}

\newcommand{\snot[1]}{{\scriptstyle \times 10}^{#1}}

 \def\norm#1{\vert #1 \vert}

 \title{On the analytical construction of halo orbits 
 and halo tubes in the elliptic restricted three-body problem}
 \author{Roc\'io I. Paez$^{1,2}$ \& Massimiliano Guzzo$^2$\\
   {\small $^1$School of Computer Science and Information Technology,
     University College Cork UCC, Cork, Ireland}\\
   {\small $^2$Dipartimento di Matematica ``Tullio Levi-Civita'',
     Universit\`a degli Studi di Padova}\\
   {\small Padova 35122 (PD) Italia}}
\date{\today}

\begin{document}
\maketitle

\begin{abstract}
  The halo orbits of the spatial circular restricted three-body
  problem are largely considered in space-flight dynamics to design
  low-energy transfers between celestial bodies. A very efficient
  analytical method for the computation of halo orbits, and the related
  transfers, has been obtained from the high-order
  resonant Birkhoff normal forms defined at the Lagrangian points L1-L2.  In
  this paper, by implementing a non-linear Floquet-Birkhoff
  resonant normal form, we provide the definition of orbits,
  as well as their manifold tubes, which exist in a large
  order approximation of the elliptic three-body problem 
  and generalize the halo orbits of the circular problem.
  Since the libration amplitude of such halo
  orbits is large (comparable to the distance of L1-L2 to the secondary
  body), and the Birkhoff normal forms are obtained through series
  expansions at the Lagrangian points, we provide also an error
  analysis of the method with respect to the orbits of the 
  genuine elliptic restricted three-body problem. 
\end{abstract} 

\section{Introduction}

In recent years, the scientific exploration of the vicinity of the
Lagrangian points, particularly in the Sun-Earth and Earth-Moon
systems, has been particularly intense. In particular, the computation
of trajectories which are in the manifolds asymptotic to orbits
librating close the Lagrangian points $L_1$-$L_2$ has gained a high
priority for space-flight dynamics: typically, the transfer design has
evolved from the familiar Earth-to-orbit concept to an
Earth-to-manifold strategy. The halo orbits of the spatial circular
restricted three-body problem have been largely considered to design
low-energy Earth-to-manifold
transfers~\cite{FarquharRep,Farquhar}. For example, a basic halo orbit
was incorporated into the trajectory for the International Sun Earth
Explorer-3 (ISEE-3) satellite, launched toward a Sun-Earth $L_1$ halo
orbit in 1978 (the satellite was the first to successfully reach a
libration point orbit). Since ISEE-3, several missions to Sun-Earth
libration point orbits have been accomplished. A current example is
the James Webb Space Telescope (JWST), designed for observations of
deep space in the infrared spectrum from an $L_2$ orbit.
 
\begin{center}
  \begin{figure}\label{fig:haloscrtbp}
    \includegraphics[width=1.\columnwidth]{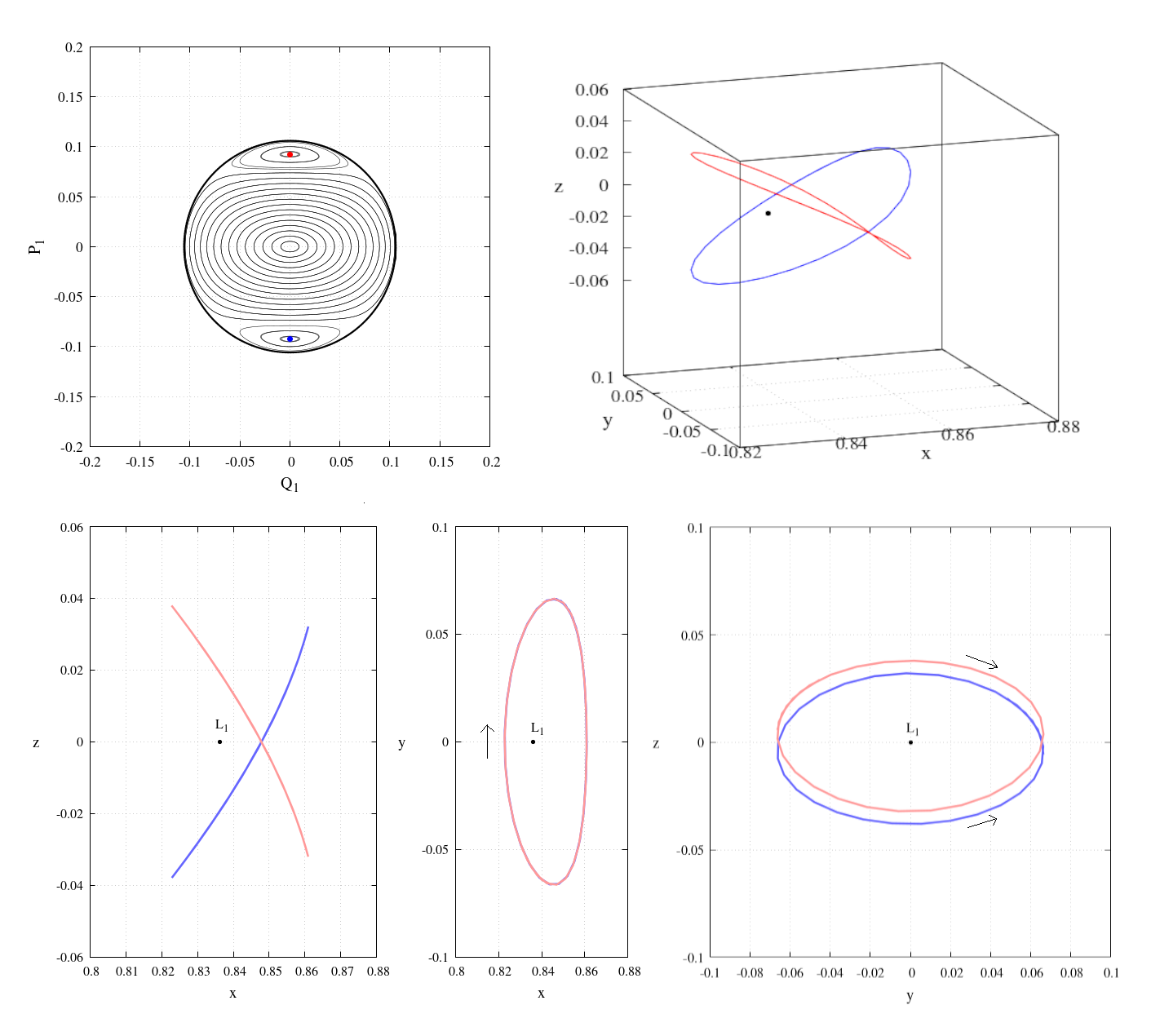}   
    \caption{\small Examples of northern (red) and southern (blue)
      halo orbits (top-right panel) computed in the CR3BP, for
      $\mu=0.0123$ (identifying the Earth-Moon system) and the Jacobi
      constant $C=3.1637151$. The top-left panel reports the phase
      portrait of the Poincar\'e section (the variables $Q_1,P_1$ on
      the section will be defined in Section 4) for the same values of
      $\mu,C$. In the bottom panels we represent the projection of
      these halo orbits on the Cartesian planes (the arrows indicate
      the sense of motion in each case).  }
  \end{figure}
\end{center}

The halo orbits are defined in the Circular Restricted Three-Body
Problem (CRTBP) from the computation of large order resonant Birkhoff
normal forms at the Lagrangian point $L_1$ or $L_2$
\cite{JM99,masdemont05,MP12,CPS13,CCP16,pucacco19}. Precisely, the
Birkhoff normal forms are used to compute analytically all the orbits
in the center manifold of the selected Lagrangian point $L_1$;
therefore, from the Poincar\'e section of the dynamics restricted to
the center manifold one defines the halo orbits.  In Figure
\ref{fig:haloscrtbp} we represent an example of the output of such a
computation: on the left panel we represent the phase portrait of the
Poincar\'e section for a sample value of the reduced
  mass $\mu$ and of the Jacobi constant $C$, and on the right panel
we represent the corresponding halo orbits in
space. We emphasize that using the Birkhoff-normal
forms, one obtains not only the orbits on the center manifold, but
also the orbits which are asymptotic to them and the orbits transiting
in its neighbourhood. Thus, they provide a complete analytic framework
to study the Earth-to-manifold transfers in the approximation of the
CRTBP. The CRTBP is indeed the main model to introduce the dynamics
close to $L_1,L_2$ of a selected secondary body (for example the
Earth).  Due to their importance for space-flight dynamics, many
efforts to use halo orbits in models more complicated than the CRTBP
have been done in the literature \cite{GMM03,HL11,LGMT13,LTBT14}. For
example, in the real Solar System the eccentricity of the orbits of
the planet identified as the secondary body, as well as the
perturbations from the other planets, limit the study to look for
orbits with features similar to the orbits identified in the
approximation of the CRTBP (see for example
\cite{LXHS13,DTT17,JJCR20,SG20,RJJC21,McCH21,SGP21}).

When we consider the short time-spans typical of space-flight dynamics
or of close encounters of a comet with a planet, the major
modification to the CRTBP is represented by the Elliptic Restricted
Three-Body Problem (ERTBP), where the orbit of the secondary body
$P_2$ performs an elliptic motion around the primary body $P_1$. The
ERTBP is conveniently represented as a non-autonomous Hamiltonian
system having the Lagrangian solutions $L_1,L_2$ but without a global
first integral, such as the Jacobi constant, which is used to define
the Poincar\'e sections and label the halo orbits in the CRTBP. In the
paper \cite{PG21} we have introduced Floquet-Birkhoff normal forms for
the ERTBP which allowed us to generalize and compute the families of
planar and vertical Lyapunov orbits generating at $L_1,L_2$, as well
as the low-energy transits from one side to the other of the secondary
body. In this paper we compute halo orbits for the ERTBP from {\it
  resonant} Birkhoff-Floquet normal forms. Despite the lack of a
global first integral for the ERTBP, the resonant Birkhoff-Floquet
normal form allows us to define a non-linear approximation of large
order of the dynamics, with an approximate local first integral, which
we call 'local energy', labelling the Poincar\'e sections close to
$L_1,L_2$. From the Poincar\'e sections we identify then the halo
orbits, which are finally mapped to the Cartesian space using a
time-dependent canonical transformation.  This method of computation
of halo orbits (as well as the method used for the CRTBP) is based on
series expansions of the Hamiltonian, truncated at a large order. The
dependence of the error on this truncation order is influenced both by
the singularity of the Hamiltonian corresponding to a collision with
$P_2$ (see \cite{PG20}) and by the well known problem of accumulation
of small divisors (see for example \cite{EGC04} and references
therein). Since the family of halo orbits forms with a minimum
libration amplitude, it is necessary to perform a test on the error
introduced with the truncation of series. 

The paper is organized as follows: in Section 2 we review some basic
properties of halo orbits in the CR3BP; in Section 3 we introduce the
resonant Floquet-Birkhoff normal forms of the ERTBP, and from these
normal forms we define the halo orbits as well as their asymptotic
manifolds; in Section 4 we illustrate an application of the method to
the Earth-Moon ERTBP, with the error analysis.

\section{Halo orbits: from the CR3BP to the ER3BP}

The CRTBP is defined by the dynamics of a particle $P$ of
infinitesimal mass attracted by two massive bodies $P_1,P_2$ revolving
around their center of mass in circular orbits. In a suitable rotating
reference frame this model admits five equilibrium points, the so
called Lagrangian points $L_1,\ldots ,L_5$, and a first integral, the
Jacobi constant, related to the energy of the particle. In this paper
we focus on the collinear Lagrangian points $L_1,L_2$, which behave,
linearly, as the product of two centers by a saddle. Due to the
center-center part, there are 4-dimensional center manifolds for $L_1$
and $L_2$, containing also periodic orbits and invariant KAM tori
\cite{Arnold-63,Kolmo-54,Moser-62}. It is convenient to consider the
3-dimensional levels of the center manifolds that we obtain when we
fix the value of the Jacobi constant; the stable and unstable
manifolds of these sets are the so called manifold tubes. The orbits
on the manifold tubes approach exponentially orbits on the center
manifold in the future (the stable manifold tubes) or in the past (the
unstable manifold tubes). From the several types of families of orbits
in the center manifolds two families are of particular interest for
Astrodynamics: the planar Lyapunov orbits and the three-dimensional
halo orbits.

The family of halo orbits results from a bifurcation in the
corresponding $L_{1}$ or $L_{2}$ Lyapunov
family~\cite{Howell,howell-01}, and extends from the vicinity of the
Lagrangian point toward the nearest massive body $P_2$.  All halo
orbits include an out-of-plane component, i.e. an amplitude component
in the (vertical) z-direction. In particular, for the $L_{1}$ halo
family, the vertical amplitude increases as the orbit moves toward
$P_2$. Because the halo family results from a pitchfork bifurcation in
the planar family, the bifurcation introduces two branches that extend
both above and below the xy-plane. A halo orbit with a maximum
out-of-plane excursion in the positive z-direction is termed a
\emph{northern} halo orbit, while the orbits with a maximum vertical
amplitude in the negative z-direction is termed \emph{southern}. A
sample northern orbit (red) with the corresponding southern halo
(blue) is plotted in Figure \ref{fig:haloscrtbp}. Note that from a
xy-projection, the direction of motion for both northern and southern
orbits about $L_{1}$ is clockwise, but when viewed from a
yz-projection, the motion of the northern orbit is clockwise while the
motion of the southern orbit is counter-clockwise.

The halo orbits of the CRTBP have been analytically computed from
computer assisted implementations of Hamiltonian perturbation theory
as well as from numerical methods 
(see for example \cite{JM99,masdemont05,CPS13,pucacco19}
for the analytic methods and \cite{Farquhar,howell-01,GM2001,QYJZ18,QGYYW19}
for the numerical ones). In this paper we consider the analytic
computations based on Hamiltonian perturbation theory, which allow not
only to compute the halo orbits, but also the orbits in their
neighbourhoods, including their stable and unstable manifolds and
transit orbits. In the CR3BP the result is achieved by computing a
resonant Birkhoff normal form of large order $N$: by neglecting the
large order remainder, one remains with and integrable Hamiltonian
system which is used to compute the Poincar\'e section of the
Hamiltonian flow on the center manifold, and consequently the halo
orbits. We extend these methods to the ERTBP by providing a definition of halo orbits in the
elliptic problem, and a method of computation based on the resonant
version of the Floquet-Birkhoff normal forms which were introduced in
\cite{PG21}. The resonant Floquet-Birkhoff normal forms will be used
to define also the manifolds tubes and the transit motions associated
to halo orbits in the ERTBP.

The ERTBP is defined by the motion of a body $P$ of infinitesimally
small mass moving in the gravity field generated by two massive bodies
$P_1$ and $P_2$, which move around their common center of mass
according to the elliptic solutions of the two-body problem. It is
convenient to represent the motion of $P$ using a rotating-pulsating
reference frame $(x,y,z)$ whose origin is in the center of mass of
$P_1$ and $P_2$, the $z$ axis is orthogonal to their motion, and the
$x,y$ axes are rotating-pulsating so that $P_1,P_2$ remain at fixed
locations on the horizontal axis $x$. With standard units of measure,
the Hamiltonian representing the motions of $P$ in this
pulsating-rotating frame is
\begin{equation}\label{eq:ori3bp}
  \begin{aligned}
    h(x,y,z,p_x,p_y,p_z,& f;e)  =  \frac{p_x^2}{2} + \frac{p_y^2}{2}
    + \frac{p_z^2}{2}- p_y \, x +
    p_x \, y \\
    & + \frac{1}{1+ e \, \cos f} \left(  \frac{1}{2}\, e \, (x^2+y^2+z^2)
    \cos f  \right. \\
    & \left. -  \frac{\mu}{\sqrt{(x-(1-\mu))^2+y^2+z^2}} -
    \frac{1-\mu}{\sqrt{(x+\mu)^2+y^2+z^2}} \right)~,
  \end{aligned}
\end{equation}
where the independent variable, denoted by $f$, corresponds to the
true anomaly of the secondary body, the parameter $\mu\in
(0,\frac{1}{2}]$ denotes the reduced mass, and $e$ denotes the
  eccentricity of the elliptic motion. The main advantage of using
  rotating--pulsating variables is that the Hamilton equations of
  (\ref{eq:ori3bp}) have five equilibrium points $L_1,\ldots ,L_5$
  located in the same positions $(x_{L_i},y_{L_i},0)$ of the
  corresponding circular problem; the collinear points $L_1,L_2$ are
  denoted by $(x,y,z,p_{x},p_y,p_z)=(x_{L_i},0,0,0,x_{L_i},0)$.
For each selected equilibrium $L_i$ we first introduce the variables
$(\mathbf{q},\mathbf{p})=(q_1,q_2,q_3,p_1,p_2,p_3)$:
\begin{equation}\label{eq:expanxL1}
  \begin{aligned}
    x & = q_1 + x_{L_i}~, \quad     &p_x & = p_1~, \\
    y & = q_2~, \quad     &p_y & = p_2 + x_{L_i}~, \\
    z & = q_3~, \quad     &p_z & = p_3~, \\
  \end{aligned}
\end{equation}
such that the equilibrium point $L_i$ is in the origin of the phase-space, 
and consider the Taylor expansion of $h$  in $(\mathbf{q},\mathbf{p})$:
\begin{equation}\label{eq:expandedH0}
H (\mathbf{q},\mathbf{p},f;e) = H_2 + H_3+ \ldots 
\end{equation}
where each term $H_j(\mathbf{q},\mathbf{p},f;e)$ is a polynomial of
degree $j$ in the variables $(\mathbf{q},\mathbf{p})$. Notice that the
zero-order term $H_0 (f;e)$ has been removed from the Hamiltonian; 
the term of order $1$ vanishes because we are expanding the
Hamiltonian at an equilibrium point; the term of second order is
{\small
\begin{equation}\label{eq:H2}
    H_2 (\mathbf{q},\mathbf{p},f;e)= \, \frac{p_1^2}{2} + \frac{p_2^2}{2} +
    \frac{p_3^2}{2}  - p_2 \, q_1 + p_1 \, q_2 
     + \, \frac{\beta\, (-2 q_1^2+q_2^2+q_3^2)}{1
    + e \cos f}  + \frac{(e \cos f)\, (q_1^2+q_2^2+q_3^2)}{2\,(1 + e \cos f)}
\end{equation}}
with
\begin{equation}\label{eq:beta}
  \beta = \frac{1}{2} \left(\frac{\mu}{\norm{1-x_{L_i} -\mu}^3} + \frac{1-\mu}{
\norm{x_{L_i}+\mu}^3} \right)~.
\end{equation}
Then, we use a combination of the Floquet theory and Birkhoff normalizations
to conjugate the Hamiltonian (\ref{eq:expandedH0}) to a normal
form which is autonomous up to a suitable large order $N$:
\begin{equation}\label{eq:BTnf}
K (\mathbf{ Q},\mathbf{ P},f;e) = K_2 (\mathbf{ Q},\mathbf{ P};e)+ K_4(\mathbf{ Q},\mathbf{ P};e)+ \ldots +K_N(\mathbf{ Q},\mathbf{ P};e)+R_{N+1}(\mathbf{ Q},\mathbf{ P},
f;e)
\end{equation}
where each term $K_j(\mathbf{ Q},\mathbf{ P};e)$ is an autonomous
polynomial of degree $j$ in the variables $(\mathbf{ Q},\mathbf{ P})$ 
and is 1-1 resonant in the sense explained below. The remainder
$R_{N+1} (\mathbf{ Q},\mathbf{ P},f;e)$ of the Taylor expansion of $K$
contains monomials from order $N+1$ and is possibly dependent on $f$.

To define the resonance properties of the polynomials $K_j(\mathbf{
  Q},\mathbf{ P};e)$ it is convenient to introduce the Birkhoff
variables $\mathbf{\hat q},\mathbf{\hat p}$ canonically
conjugated to the real variables $(\mathbf{Q},\mathbf{P})$ by the
linear transformation:
\begin{equation}\label{birkhoffcomplex}
 Q_3=\hat q_3\ \ ,\ \  P_3=\hat p_3\ \ ,\ \ 
 Q_j= \frac{{\hat q}_j+\mathrm{i}\,{\hat p}_j}{\sqrt{2}}~,\ \ 
 P_j= \frac{\mathrm{i}\,{\hat q}_j+{\hat p}_j}{\sqrt{2}}~,\,\,j=1,2  . 
\end{equation}
In this paper  we consider the resonant Floquet-Birkhoff normal forms such that
all the terms $K_j(\mathbf{Q},\mathbf{P};e)$, when represented
using the variables $\mathbf{\hat q},\mathbf{\hat p}$, 
are the sum of monomials 
$$
a_{\mathbf{m},\mathbf{l}}{\hat q}_1^{m_1}{\hat q}_2^{m_2}{\hat q}_3^{m_3}{\hat p}_1^{l_1}
{\hat p}_2^{l_2}{\hat p}_3^{l_3}
$$
with $m_3=l_3$ and $(l_1-m_1)+(l_2-m_2)=0$. This means that, if we introduce
the action-angle variables $I_1,I_2,\theta_1,\theta_2$ for the elliptic
motions and the hyperbolic variables $I_3,\theta_3$ such that:
\begin{equation}
  \begin{aligned}
   \hat q_1 & = - \mathrm{i} \, \sqrt{I_1} \, \mathrm{e}^{\mathrm{i}\, \theta_1}~,
    \qquad
    &  \hat p_1 & = \sqrt{I_1} \, \mathrm{e}^{-\mathrm{i}\, \theta_1}~, \\
  \hat     q_2 & = - \mathrm{i} \, \sqrt{I_2} \, \mathrm{e}^{\mathrm{i}\, \theta_2}~,
    \qquad
    &  \hat p_2 & = \sqrt{I_2} \, \mathrm{e}^{-\mathrm{i}\, \theta_2}~,  \\
  \hat     q_3 & =\sqrt{I_3} \, \mathrm{e}^{\theta_3}~,
    \qquad
    &  \hat p_3 & = \sqrt{I_3} \, \mathrm{e}^{- \theta_3}~, 
  \end{aligned}\label{actionangle1}
\end{equation}
the terms $K_j$ are independent of $\theta_3$, and depend on $\theta_1,\theta_2$
only through the resonant combination $\theta_1-\theta_2$. 
\vskip 0.4 cm
\noindent
{\bf Remark.} In our paper~\cite{PG21} we constructed non-resonant
Floquet-Birkhoff normal forms, so that all the terms
$K_j(\mathbf{Q},\mathbf{P};e)$ were integrable in the sense that, when
represented using the variables $\mathbf{\hat q},\mathbf{\hat p}$,
they depended on the variables only through the combinations $i\hat
q_1\hat p_1$, $i\hat q_2\hat p_2$ (the actions of the elliptic motions
expressed in Birkhoff complex variables ) and $\hat q_3\hat p_3$ (the
action of the hyperbolic motion). This type of normal form, which is
specifically designed for the efficient analytic computation of the
planar and vertical Lyapunov orbits, as well as their manifold tubes,
can be constructed if the three frequencies describing the motion are
strictly non resonant up to order $N$. As pointed in~\cite{pucacco19},
while for the Earth-Moon system ($\mu=0.0123$) indeed no exact
low order resonance
takes place, the linear frequencies lay very close
to a 1-1 resonance opening the door, for the CRTBP, to the appearance
of the halo orbits at suitable large values of the Hamiltonian.
Therefore, while very efficient in the context of planar Lyapunov
orbits and their manifold tubes, the construction presented
in~\cite{PG21} obviously excludes the computation of Halo orbits. 
\vskip 0.4 cm
Let us now define the halo orbits in the ERTBP using the resonant
Floquet-Birkhoff normal forms (\ref{eq:BTnf}). Since the dependence on
$f$ is relegated within the remainder of the normal form, the
definition of the halo orbits as well as of their manifolds tubes in
the normal form variables $(\mathbf{Q},\mathbf{P})$ is obtained as in the CRTBP from
the approximated Hamiltonian:
\begin{equation}\label{eq:BTnfintegable}
{\cal K}(\mathbf{Q},\mathbf{P};e)=K_2 (\mathbf{Q},\mathbf{P};e)+ K_4(\mathbf{Q},\mathbf{P};e)+ \ldots +K_N(\mathbf{Q},\mathbf{P};e) , 
\end{equation}
which we call {\it local energy}. Since the Hamiltonian
${\cal K}$ is integrable
by quadratures, we can compute and classify all the solutions
of its Hamilton equations. In particular:
\begin{itemize}
\item[-] for $ Q_3,P_3=0$ we have the center manifold ${\cal M}$
  of the equilibrium $(\mathbf{Q},\mathbf{P})=(0,\ldots ,0)$
  for the Hamiltonian flow of ${\cal K}$; we denote by ${\cal M}_\kappa$
  the intersection of the center manifold ${\cal M}$ with the level
  set ${\cal K}(\mathbf{Q},\mathbf{P};e)=\kappa$ of the local energy;
\item[-] for $Q_3=0,P_3\ne 0$ and for  $Q_3\ne 0,P_3= 0$ 
we have the local stable and unstable manifolds
  of ${\cal M}$;
\item[-] for suitably large values of $\kappa$ we have the two
  periodic orbits of ${\cal M}_\kappa$ which are identified as halo
  orbits;
\item[-] for initial conditions close to the manifold tubes
  of the halo orbits we find orbits which approach the
  halo orbits from one side and then transit to the other
  side  with a fly-by with the halo orbit (which we call the halo transit
  orbits),
  as well orbits which approach the  halo orbits from one side and then bounce back.
  \end{itemize}
Finally, the orbits found in the normal-form variables 
are mapped to the original Cartesian
variables with the $f$-dependent canonical transformation
\begin{displaymath}\label{calX}
(\mathbf{q},\mathbf{p})={\cal X}(\mathbf{Q},\mathbf{P};f)
\end{displaymath}
conjugating the Hamiltonian (\ref{eq:expandedH0}) to the normal form
(\ref{eq:BTnf}).  
We remark that in the space of the Cartesian
variables the halo orbits are transformed by ${\cal X}$ to
quasi-periodic orbits; we call halo torus the set of all these periodic orbits. 

\section{Construction of the resonant Floquet-Birkhoff normal form}

The resonant Floquet-Birkhoff normal form, as well as the canonical
transformation from the Cartesian variables to the normal form variables,
are represented as Taylor-Floquet expansions of terms
proportional to
$$
e^{i\nu f}{\hat q}_1^{m_1}{\hat q}_2^{m_2}{\hat q}_3^{m_3}{\hat p}_1^{l_1}{\hat p}_2^{l_2}{\hat p}_3^{l_3}
$$
up to truncation orders ${\cal N}_1,{\cal N}_2$
for the polynomial variables and for the true anomaly:
$$
m_1+m_2+m_3+l_1+l_2+l_3 \leq {\cal N}_1\ \ ,\ \ \norm{\nu}\leq {\cal N}_2 ~.
$$
The coefficients of all these terms are represented in floating point
numbers, obtained from an algebraic manipulator program
performing the transformation to the normal form
variables as the composition of:
\begin{itemize}
\item[(i)] A canonical Floquet transformation:
\begin{displaymath}
(\mathbf{q},\mathbf{p})={\cal C}(f;e)(\mathbf{\tilde q},\mathbf{\tilde p})
\end{displaymath}
conjugating the Hamiltonian (\ref{eq:expandedH0}) to an Hamiltonian:
\begin{equation}\label{eq:expandedHt}
\tilde H (\mathbf{\tilde q},\mathbf{\tilde p},f;e) = \tilde H_2
(\mathbf{\tilde q},\mathbf{\tilde p};e)+ \tilde H_3(\mathbf{\tilde
  q},\mathbf{\tilde p},f;e)+ \ldots
\end{equation}
where each term $ \tilde H_j(\mathbf{\tilde q},\mathbf{\tilde p},f;e)$
is polynomial of degree $j$ in the variables
$\mathbf{\tilde  q},\mathbf{\tilde p}$ and periodic in
$f$ with period $2\pi$, while
$\tilde H_2 (\mathbf{\tilde q},\mathbf{\tilde p};e)$ is autonomous.

As it is well known, the Floquet transformation is not unique, since
its definition depends on the arbitrary choice of a logarithm of the
monodromy matrix associated to the equations of motion linearized at
the Lagrange equilibrium.  As pointed out in \cite{PG21}, the
subsequent Birkhoff normalizations of Hamiltonian
(\ref{eq:expandedHt}) perform much better if among all the possible
Floquet transformations of (\ref{eq:expandedH0}) there is one which is
close to the identity.  Also in this paper we define the Floquet
transformation for the ERTBP by selecting a close to the identity one,
as shown in \cite{PG21}.

\item[(ii)] A linear canonical transformation:
\begin{equation}\label{eq:linD2}
 ( \mathbf{\tilde q}, \mathbf{\tilde p}) = {\cal D}(\mathbf{\hat q},\mathbf{\hat p})
\end{equation}
giving $\tilde H_2 (\mathbf{\tilde q},\mathbf{\tilde p};e)$ the normal form:
\begin{equation}\label{eq:k2}
\hat K_2( \mathbf{\hat q},\mathbf{\hat p})= \sigma_1{ {\hat q}_1^2+{\hat
    p}_1^2\over 2}+\sigma_2{ {\hat q}_2^2+{\hat p}_2^2\over 2}+\lambda
\hat q_3 \hat p_3 .
\end{equation}
We denote by $\hat K_j(\mathbf{\hat q},\mathbf{\hat p},f;e)$ the 
image
of all the other polynomials $\hat K_j(\mathbf{\hat q},\mathbf{\hat
  p},f;e)= \tilde H_j ( {\cal D}(\mathbf{\hat q}, \mathbf{\hat
  p}),f;e)$.

\item[(iii)] A sequence of $N-2$ Birkhoff
  transformations giving the Hamiltonian the final resonant normal form
  (\ref{eq:BTnf}).
\end{itemize}

The Floquet transformation (i) and the linear transformation ${\cal
  D}$ are discussed in~\cite{PG21}, and require no modifications to
adapt to the resonant case. Therefore we provide in this paper all the
details of the resonant Birkhoff transformations (iii). The two linear
transformations (i) and (ii) conjugate the Hamiltonian
(\ref{eq:expandedH0}) to
\begin{equation}\label{eq:expham}
  \hat{H}(\hat{\mathbf{q}},\hat{\mathbf{p}},F,f) = F + 
\hat H_2(\hat{\mathbf{q}},\hat{\mathbf{p}}) + \sum_{j\geq 3} \hat{H}_j(\hat{\mathbf{q}},\hat{\mathbf{p}},f;e)~,
\end{equation}
where the variable $F$, conjugated to $f$, has been introduced in
order to conveniently deal with an autonomous Hamiltonian and the
terms $\hat{H}_j$ for $j\geq 3$ are polynomials of degree $j$ in the
variables $\mathbf{\hat q},\mathbf{\hat p}$ and periodic in $f$ with
period $2\pi$. The terms $\hat{H}_j$ with $j\geq 3$ are represented as
sum of monomials of the form
\begin{equation}\label{eq:genmono}
  a^{(j)}_{\mathbf{m},\mathbf{l},\nu} {\mathrm e}^{i\nu f}\,
  \hat{q}_1^{m_1} \hat{q}_2^{m_2} \hat{q}_3^{m_3} \hat{p}_1^{l_1}
  \hat{p}_2^{l_2} \hat{p}_3^{l_3}~,~~\sum_{i=1}^3 (m_i+l_i) = j~.
\end{equation}
The objective of the resonant Birkhoff transformations is to tackle in
a single algorithmic procedure two different effects: to remove the
explicit dependence of $\hat{H}$ on $f$ up to a large finite order
$N$, and to define a normal form Hamiltonian which can be exploited to
study the resonance generating the halo orbits. This is achieved with
a close to the identity canonical transformation ${\cal C}_N$
conjugating the Hamiltonian \eqref{eq:expham}, that now we denote as
the initial Hamiltonian $\hat{H}^{(2)}$, to a normal form Hamiltonian
\begin{equation}\label{eq:minormf}
\hat{H}^{(N)} = F + \sum_{j=2}^N K^{(N)}_j(\hat{\mathbf{q}},\hat{\mathbf{p}})+
\sum_{j\geq N+1} \hat{H}^{(N)}_j (\hat{\mathbf{q}},\hat{\mathbf{p}},f)
\end{equation}
where the functions $K^{(N)}_j$ do not depend on $F,f$ and are 
polynomials of degree $j$ depending on $\hat{\mathbf{q}},\hat{\mathbf{p}}$
only through monomials of the form:
\begin{equation}\label{eq:genmonoK}
  a^{(j,N)}_{\mathbf{m}} 
  \hat{q}_1^{m_1} \hat{q}_2^{m_2} \hat{q}_3^{m_3} \hat{p}_1^{l_1}
  \hat{p}_2^{l_2} \hat{p}_3^{l_3}~,
\end{equation}
with:
\begin{equation}\label{eq:genmonoKl}
  (l_1-m_1)+(l_2-m_2) = 0 \,\, \ \ \mathrm{and}\ \  \,\, l_3 = m_3, 
  \,\, \sum_{i=1}^3  m_i + l_i = j~.
\end{equation}
The remainder terms $\hat{H}^{(N)}_j$ are polynomials of degree 
$j$ with coefficients depending periodically on $f$ with period $2\pi$, 
represented as sum of monomials of the form
\begin{equation}\label{eq:genmonoN}
    a^{(j,N)}_{\mathbf{m},\mathbf{l},\nu}
  {\mathrm e}^{i\nu f}\, \hat{q}_1^{m_1} \hat{q}_2^{m_2}
  \hat{q}_3^{m_3} \hat{p}_1^{l_1} \hat{p}_2^{l_2}
  \hat{p}_3^{l_3}~,\ \ \ \ ~~\sum_{i=1}^3 (m_i+l_i) = j~.
\end{equation}
The canonical transformation ${\cal C}_N$ is constructed from
the composition of a sequence of 
$N-2$ elementary canonical Birkhoff transformations. Precisely,
we define the sequence of canonical transformations:
\begin{equation}\label{eq:totalC}
  {\cal C}_J= {\hat {\cal C}}_{{\chi}_{_J}}\, \circ \, {\cal C}_{J-1}~, \ \ J=3,\ldots , N
\end{equation}
conjugating $\hat{H}:=\hat{H}^{(2)}$ to the intermediate
Floquet-Birkhoff normal form Hamiltonians:
\begin{equation}\label{eq:intermediateminormf}
\hat{H}^{(J)} := \hat{H}^{(J-1)}\circ {\cal C}_J =
 F + \sum_{j=2}^J K^{(J)}_j(\hat{\mathbf{q}},\hat{\mathbf{p}})+
\sum_{j\geq J+1} \hat{H}^{(J)}_j (\hat{\mathbf{q}},\hat{\mathbf{p}},f)
\end{equation}
with the property that $K^{(J)}_j$ do not depend on $F,f$ and are
polynomials of degree $j$ depending on
$\hat{\mathbf{q}},\hat{\mathbf{p}}$ only through
monomials of the form (\ref{eq:genmonoK}) with powers satisfying
(\ref{eq:genmonoKl}), while $\hat{H}^{(J)}_j$ are polynomials of degree $j$ with
coefficients depending periodically on $f$ with period $2\pi$.

The transformations are defined as follows: ${\cal C}_{2}$ is the
identity while $\hat{\cal C}_{\chi_{_J}}$ is the Hamiltonian flow at
time $f=1$ of generating functions $\chi_{_J}$ defined from the
coefficients of $\hat{H}^{(J-1)}$. Below we describe the definition of
the generating functions $\chi_{_J}$ and the steps required for the
algorithmic computation of each canonical transformation ${\cal C}_N$
and Hamiltonian $\hat{H}^{(N)}$ using the Lie series method (for an
introduction to the method, see~\cite{LaPlata,Pisa}) and implemented
with a computer algebra system in the examples presented in this
paper. For each $J\geq 3$ we assume that the Hamiltonian
$\hat{H}^{(J-1)}$ and the canonical transformation ${\cal C}_{N-1}$
are known, and we proceed as follows.
\vskip 0.2 cm
\noindent
First, from $\hat{H}^{(J-1)}$ we compute the generating function
${\chi_{_J}}$:
  \begin{equation}\label{eq:genfunc-a}
  \chi_{_J} = \hspace{-0.6cm}
  \sum_{\substack{(\mathbf{m}, \mathbf{l}, \nu)\in {\cal L}_J}}
  \hspace{-0.3cm}
    \frac{-a^{(J-1)}_{\nu,m_1,m_2,m_3,l_1,l_2,l_3}}{
      \mathrm{i}\,\sigma_1(l_1-m_1) + \mathrm{i}\,\sigma_2(l_2-m_2)
      + \lambda \, (l_3-m_3) + \mathrm{i}\, \nu} \,
    {\mathrm e}^{i\nu f}\,
    \hat{q}_1^{m_1} \hat{q}_2^{m_2} \hat{q}_3^{m_3} \hat{p}_1^{l_1}
    \hat{p}_2^{l_2} \hat{p}_3^{l_3}
\end{equation}
where
\begin{displaymath}
{\cal L}_{_J}=\left\{ (\mathbf{m},\mathbf{l},\nu)\in {\Bbb N}^3\times {\Bbb N}^3 \times {\Bbb Z}:  \sum_{j=1}^3
(l_j+m_j)=J,\ {\rm and}\
\norm{l_1-m_1+l_2-m_2}
+ \norm{l_3 - m_3}+ \norm{\nu}\geq 1 \right\}~.
\end{displaymath}    
Next, we compute  explicitly  the canonical transformation
\begin{displaymath}
  \hat{\cal C}_{\chi_{_J}}
  (\hat{\mathbf{q}}^{(J)},\hat{\mathbf{p}}^{(J)},F^{(J)},f^{(J)})=
(\hat{\mathbf{q}}^{(J-1)},\hat{\mathbf{p}}^{(J-1)},F^{(J-1)},f^{(J-1)})~,
\end{displaymath}
as the Lie series
\begin{equation}\label{eq:theCchin}
 \zeta = e^{\,L_{\chi_{_J}}} \zeta' := \zeta' +
 \{\zeta', \chi_{_J} \}+ {1\over 2} \{ \{\zeta',
 \chi_{_J} \}, \chi_{_J} \}+\ldots ~,
\end{equation}
where $L_{\chi_{_J}} := \{\cdot, \chi_{_J} \}$, and $\zeta, \zeta'$ denote
any couple of variables
$\hat{\mathbf{q}}^{(J-1)},\hat{\mathbf{q}}^{(J)}$,
$\hat{\mathbf{p}}^{(J-1)},\hat{\mathbf{p}}^{(J)}$ or
$F^{(J-1)},F^{(J)}$.  The transformed Hamiltonian is
  computed as a Lie series as well:
\begin{equation}\label{eq:newham}
  \hat{H}^{(J)} = \hat {\cal
      C}_{\chi_{_J}}\, \hat{H}^{(J-1)}=
  e^{L_{\chi_{_J}}}\, \hat{H}^{(J-1)} .
\end{equation}
The iteration ends for $J=N$, and finally, by reintroducing real
canonical variables,
\begin{equation}\label{eq:fromqtoQ}
  \begin{aligned}
  \hat{q}_1^{(N)} &= \frac{Q_1 - \mathrm{i}\, P_1}{\sqrt{2}}~, \quad
  &\hat{p}_1^{(N)} &= \frac{P_1 - \mathrm{i}\, Q_1}{\sqrt{2}}~,\\
  \hat{q}_2^{(N)} &=  \frac{Q_2 - \mathrm{i}\, P_2}{\sqrt{2}}~, \quad
  &\hat{p}_2^{(N)} &=\frac{P_2 - \mathrm{i}\, Q_2}{\sqrt{2}}~,\\
  \hat{q}_3^{(N)} &= Q_3~, \quad
 &\hat{p}_3^{(N)} &= P_3~,
  \end{aligned}
\end{equation}
and by suitably identifying the terms $\hat K_j$ with $K_j$, and
disregarding the dummy action $F^{(N)}$, we recover the final
Floquet-Birkhoff normal form as in Eq.~\eqref{eq:BTnf}.

\section{Resonant dynamics in the center manifold of the elliptic Earth-Moon system}

In this Section we illustrate the use of the resonant Floquet-Birkhoff
normal forms to represent the dynamics related to the halo orbits in a
model problem, which we identify as the Earth-Moon ERTBP. The
relevance of the Earth-Moon halo orbits for the space-flight dynamics
has been considered in several papers
\cite{GJMS,masdemont05,KLMR06,ZCMD}. The basic model to study the
dynamics of a spacecraft in the Earth-Moon system is the Circular
Restricted Three-Body Problem (CRTBP) with the Earth and the Moon as
primaries.  This model, although simplistic compared to the model of
the Solar System which is considered to compute modern ephemerides,
had nevertheless provided deep insights regarding the dynamics of
small bodies in the Solar System.

For the value of $\mu = 0.0123$ and $e = 0.0549006$ considered in the
present paper (representing the Earth-Moon ERTBP), after 4
normalization steps ($N=6$), the normal form
$K(\mathbf{Q},\mathbf{P};e)$ introduced in \eqref{eq:BTnf} takes the form:
\begin{displaymath}
  K(\mathbf{Q},\mathbf{P};e) = K_2(\mathbf{Q},\mathbf{P};e) + K_4(\mathbf{Q},\mathbf{P};e)
  + K_6(\mathbf{Q},\mathbf{P};e)+ R_7(\mathbf{Q},\mathbf{P},f;e)
\end{displaymath}
where
\begin{displaymath}
  \begin{aligned}
    K_2(\mathbf{Q},\mathbf{P};e)=&
      2.33662 \, \frac{Q_1^2+P_1^2}{2}
    + 2.27111 \, \, \frac{Q_2^2+P_2^2}{2}
    + 2.93590 \, Q_3\,P_3~, \\
    K_4(\mathbf{Q},\mathbf{P};e)=&
    - 1.76908 \, P_1^4
    + 1.74292 \, P_1^2 \, P_2^2    
    - 1.58163 \, P_2^4 
    - 3.53816 \, P_1^2 \, Q_1^2\\
    & \,
    - 3.33654 \, P_2^2 \, Q_1^2
    - 1.76908 \, Q_1^4 
    + 10.15894 \, P_1 \, P_2 \, Q_1 \, Q_2
    - 3.33654 \, P_1^2 \, Q_2^2 \\
    & \,
    - 3.16326 \, P_2^2 \, Q_2^2   
    + 1.74292 \, Q_1^2 \, Q_2^2
    - 1.58163 \, Q_2^4 
    - 16.44122 \, P_1^2 \, P_3 \, Q_3 \\
    & \,
    - 15.03657 \, P_2^2 \, P_3 \, Q_3
    - 16.44122 \, Q_1^2 \, P_3 \, Q_3 
    - 15.03657 \, Q_2^2 \, P_3 \, Q_3
    - 9.57863 \, P_3^2 \, Q_3^2 \\    
    K_6(\mathbf{Q},\mathbf{P};e)= &
    - 3.13968 \, P_1^6 \,
    + 6.82217 \, P_1^4 \, P_2^2
    + 5.97758 \, P_1^2 \, P_2^4
    - 1.99158 \, P_2^6 \\
    & \,
    - 9.41904 \, P_1^4 \, Q_1^2
    - 1.38474 \, P_1^2 \, P_2^2 \, Q_1^2
    - 7.06430 \, P_2^4 \, Q_1^2
    - 9.41905 \, P_1^2 \, Q_1^4\\
    & \,
    - 8.20691 \, P_2^2 \, Q_1^4
    - 3.13968 \, Q_1^6
    + 30.05816 \, P_1^3 \, P_2 \, Q_1 \, Q_2
    + 26.08374 \, P_1 \, P_2^3 \, Q_1 \, Q_2\\
    & \,
    + 30.05816 \, P_1 \, P_2 \, Q_1^3 \, Q_2
    - 8.20690 \, P_1^4 \, Q_2^2 
    - 1.08671 \, P_1^2 \, P_2^2 \, Q_2^2
    - 5.97474 \, P_2^4 \, Q_2^2\\
    & \,
    - 1.38473 \, P_1^2 \, Q_1^2 \, Q_2^2
    - 1.08671 \, P_2^2 \, Q_1^2 \, Q_2^2
    + 6.82217 \, Q_1^4 \, Q_2^2
    + 26.08374 \, P_1 \, P_2 \, Q_1 \, Q_2^3\\
    & \,
    - 7.06430 \, P_1^2 \, Q_2^4
    - 5.97474 \, P_2^2 \, Q_2^4
    + 5.97758 \, Q_1^2 \, Q_2^4
    - 1.99158 \, Q_2^6\\
    & \,
    - 11.98956 \, P_1^4 \, P_3 \, Q_3
    - 96.12101 \, P_1^2 \, P_2^2 \, P_3 \, Q_3
    - 3.55055 \, P_2^4 \, P_3 \, Q_3\\
    & \,
    - 23.97913 \, P_1^2 \, P_3 \, Q_1^2 \, Q_3
    - 15.47041 \, P_2^2 \, P_3 \, Q_1^2 \, Q_3
    - 11.98956 \, P_3 \, Q_1^4 \, Q_3\\
    & \,
    - 161.30118 \, P_1 \, P_2 \, P_3 \, Q_1 \, Q_2 \, Q_3
    - 15.47041 \, P_1^2 \, P_3 \, Q_2^2 \, Q_3
    - 7.10110 \, P_2^2 \, P_3 \, Q_2^2 \, Q_3\\
    & \,
    - 96.12101 \, P_3 \, Q_1^2 \, Q_2^2 \, Q_3
    - 3.55055 \, P_3 \, Q_2^4 \, Q_3
    - 105.42195 \, P_1^2 \, P_3^2 \, Q_3^2\\
    & \,
    - 70.52337 \, P_2^2 \, P_3^2 \, Q_3^2
    - 105.42195 \, P_3^3 \, Q_1^2 \, Q_3^2
    - 70.52337 \, P_3^2 \, Q_2^2 \, Q_3^2\\
    & \,
    - 54.46117 \, P_3^3 \, Q_3^3
    \end{aligned}
\end{displaymath}
We exploit the tailored constructed resonant normal form in order to
compute the halo orbits, for a certain value of the local
energy. First, we analyze the dynamics on the center manifold by
computing its Poincar\'e surfaces of section. We denote by $\hat
K_{CM}(Q_1,Q_2,P_1,P_2;e)$ the 2-degrees of freedom Hamiltonian of the
system restricted to the center manifold, expressed with the real
canonical variables $\mathbf{Q},\mathbf{P}$ (as a matter of fact, only
$Q_1,Q_2,P_1,P_2$ are needed):
\begin{eqnarray}\label{eq:Kcm}
  K_{CM}(Q_1,Q_2,P_1,P_2;e) &:=&
  K(Q_1,Q_2,0,P_1,P_2,0;e)\cr
  &:=&  K_{CM,2}(Q_1,Q_2,P_1,P_2;e)+
  K_{CM,4}(Q_1,Q_2,P_1,P_2;e)  + ...
\end{eqnarray}
For fixed values $\kappa$ of the local energy, we consider the
Poincar\'e section defined by the flow of $K_{CM}$ and the
surface:
\begin{equation}\label{pssd}
\Sigma_\kappa = \{ (Q_1,Q_2,P_1,P_2):\ \ K_{CM}(Q_1,Q_2,P_1,P_2;e)=\kappa\ \ ,\ \ 
Q_2=0,\ \ P_2>0\} ,
\end{equation}
which is parameterized by the variables $Q_1,P_1$. 
\begin{figure}
\centering
\includegraphics[width=1.\columnwidth]{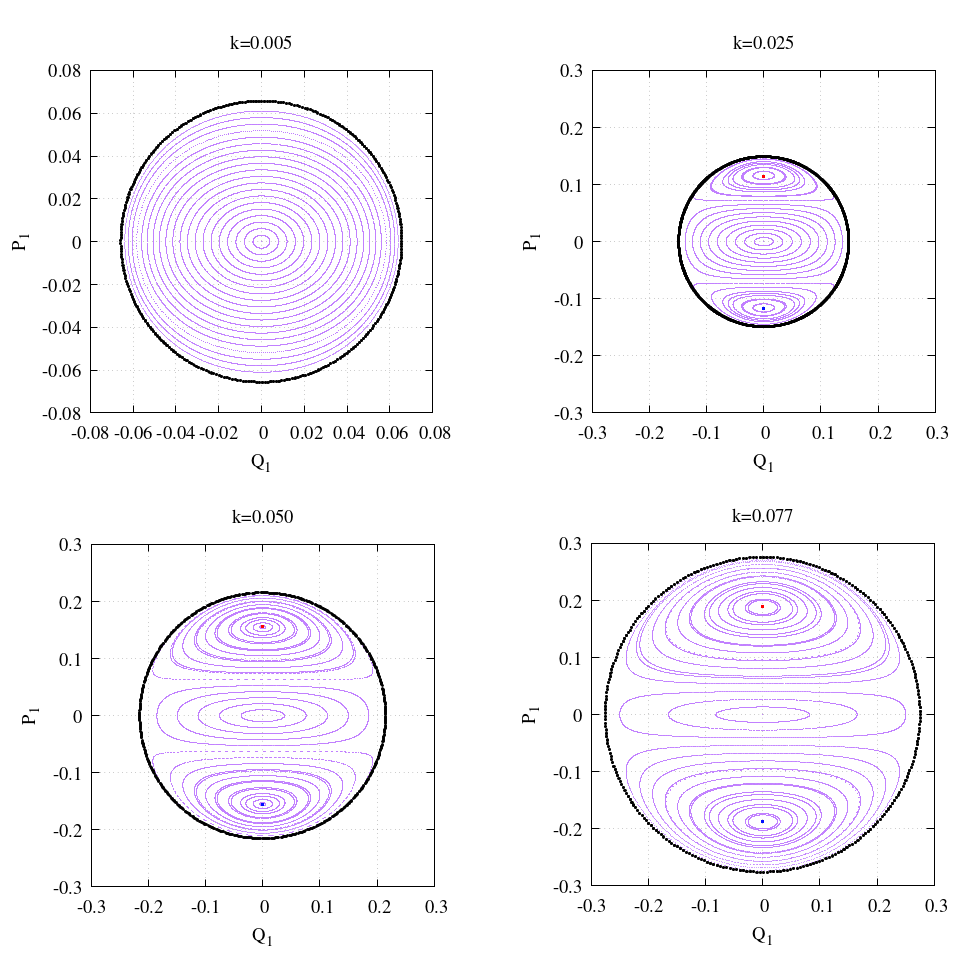}
\caption{Representation of the
  Poincar\'e sections of the flow on the center
  manifold originating at the Lagrangian solution $L_1$ of 
  the Earth-Moon system, for a sample of values of the  local energy $\kappa$,
 in the plane of the normal form variables $Q_1,P_1$.}\label{fig:ex_pss}
\end{figure}
In figure \ref{fig:ex_pss} we represent a sample of these Poincar\'e
sections numerically computed for increasing values of the local
energy $\kappa = 0.005$, $0.025$,$0.050$, $0.077$; in figure
\ref{fig:cartplots} we represent these Poincar\'e sections in the
space of the Cartesian variables, via the transformation (\ref{calX})
computed for $Q_2,Q_3,P_3,f=0$. In the phase-portraits of the
Poincar\'e sections we identify the following families of peculiar
motions\footnote{The following description refers to the flow which is
  obtained from the Hamiltonian of the ERTBP by neglecting the
  remainder $R_7$ in the Floquet-Birkhoff normal
  form. When considering the non approximated flow of
    the elliptic restricted three-body problem, the description is
    affected by an error which is discussed in Sections 4.1 and 4.2.}
\vskip 0.2 cm
\noindent
\begin{itemize}
\item[-]{\bf Vertical Lyapunov tori:} Since $\hat
K_2,\hat K_4,\hat K_6$ do not contain monomials with $l_1=1,m_1=0$ or
$l_1=0,m_1=1$, the origin $(Q_1,P_1)=(0,0)$ is a fixed point of all the
Poincar\'e sections, corresponding to a periodic orbit in the family of
the vertical Lyapunov orbits. Since the canonical transformation
${\cal X}(\mathbf Q,\mathbf P,f;e)$ maps these periodic orbits to tori
of the Cartesian space, the origin of the Poincar\'e sections provides
the family of vertical Lyapunov tori.

\item[-]{\bf Planar Lyapunov tori:} The borders of the
Poincar\'e sections (which, strictly speaking, do not belong to the
section $\Sigma_\kappa$), which are obtained for the limit
initial conditions $(Q_2,P_2)=0$ and $Q_1,P_1$ satisfying:
$$
K_{CM}(Q_1,0,P_1,0;e)=\kappa  ,
$$
correspond to the family of the planar Lyapunov orbits, which are
mapped to the planar Lyapunov tori of the Cartesian space. In fact,
since $\hat K_2,\hat K_4,\hat K_6$ do not contains monomials with
$l_2=1,m_2=0,l_3,m_3=0$ or $l_2=0,m_2=1,l_3,m_3=0$, each solution of
the Hamilton equations of $K_{CM}$ with $(Q_2(0),P_2(0))=(0,0)$,
satisfies $(Q_2(t),P_2(t))=(0,0)$ for all $t$, thus providing a planar
periodic orbit. Since the canonical transformation ${\cal X}(\mathbf
Q,\mathbf P,f;e)$ maps these orbits to tori of the Cartesian space,
the limit border of the Poincar\'e section provides
the family of planar Lyapunov tori. We remark that
both the planar and vertical Lyapunov tori are more efficiently
computed with the non-resonant normal forms defined in the paper
~\cite{PG21}.

\item[-]{\bf Halo tori.} We identify the halo orbits of the ERTBP as
  the fixed points of the Poincar\'e section of the Hamiltonian system
  defined by $K_{CM}$ (which appear in addition to the central one
  identified by $(Q_1,P_1) = (0,0)$) for all the larger values of the
  local energy $\kappa = 0.025$,$0.050$, $0.077$. As for the CRTBP
  (see \cite{MP12,CCP16,pucacco19}), the halo orbits are better
  described by introducing normal form variables which are adapted to
  the 1-1 resonance. First, we introduce on the center manifold the
  action-angle variables $\theta_1,\theta_2,I_1,I_2$ defined in
  (\ref{actionangle1}), and then the action-angle variables
  $\phi,\chi,J_\phi,J_\chi$ adapted to the 1-1 resonance defined by:
\begin{equation}\label{eq:resvar2}
  \begin{aligned}
    \theta_1 & =\phi + \chi ~, \quad \quad &\theta_2 & = \chi \\
    I_1 &= J_\phi~, \quad \quad &I_2 &= J_\chi - J_\phi~.
  \end{aligned}
\end{equation}
Since the representation of the Hamilton function $K_{CM}$ in the
action-angle variables does not depend on the angle $\chi$, the
conjugate action $J_\chi$ is a first integral and the motion of the couple
$\phi,J_\phi$ is computed from the 1-degree of freedom
reduced Hamiltonian system obtained for fixed values of $J_\chi$. The
halo orbits are computed as the equilibrium points of the reduced system, and 
correspond to fixed points of the Poincar\'e section with $Q_1=0,P_1\ne 0$.
Because the action-angle variables $\phi,J_\phi$ are singular for $J_\phi=0$,
the computation of the equilibrium points of the reduced system is better
performed using the non-singular canonical variables:
\begin{equation}
\tilde x = \sqrt{2\, J_\phi} \sin \phi~, \qquad \tilde y = \sqrt{2\, J_\phi} \cos \phi~.
\end{equation}
Therefore, by computing $K_{CM}$ in the variables $\chi,J_\chi,\tilde x,\tilde y$ we obtain a function:
$$
\tilde K_{CM} := \tilde K_{CM}(J_\chi,\tilde x,\tilde y)
$$
which is a polynomial of order 6 in the variables $\tilde x,\tilde y$,
with coefficients depending polynomially on $J_\chi$ (up to order
3). We therefore proceed by computing the equilibrium points of the
reduced Hamiltonian $\tilde K_{CM}$ with $\tilde x=0, \tilde y\ne 0$,
and the corresponding initial conditions on the Poincar\'e section
obtained for $\chi=0$. Finally, we remark that the equilibrium points
of the reduced system provide periodic orbits in the center manifold,
which are projected to families of quasi-periodic orbits of the
Cartesian space, which we call the family of halo tori of the ERTBP.
\end{itemize}

\begin{figure}
  \centering
   \includegraphics[width=0.5\columnwidth]{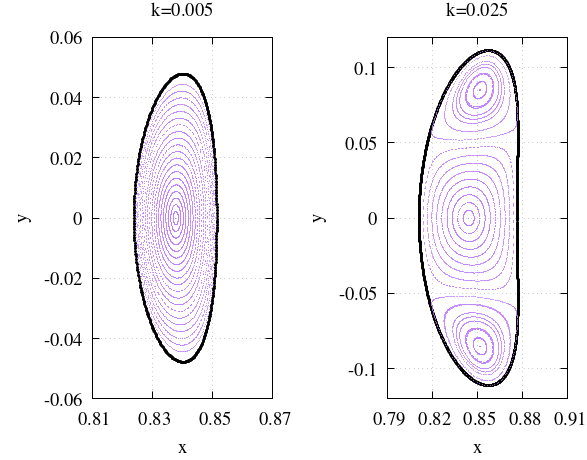}\includegraphics[width=0.5\columnwidth]{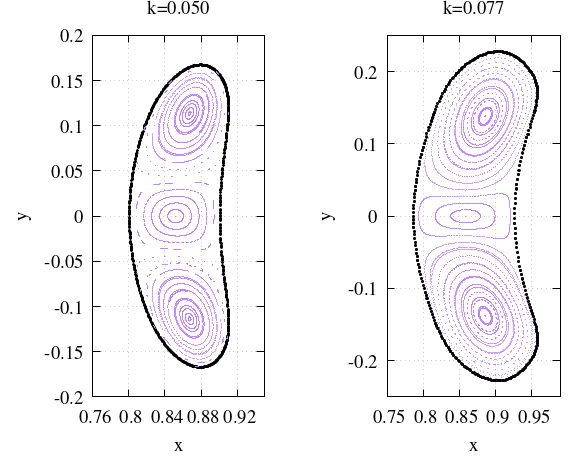}  
  \caption{Representation of the same Poincar\'e sections of 
    fig. \ref{fig:ex_pss} in the plane of the Cartesian coordinates $x,y$,
    computed by transforming the values of the normal form variables for $f=0$.
    The thick black curve
  corresponds to the section $f=0$ of the planar Lyapunov torus.}\label{fig:cartplots}
\end{figure}

\begin{figure}[h]\label{twohalos}
  \centering
 \includegraphics[width=0.5\columnwidth]{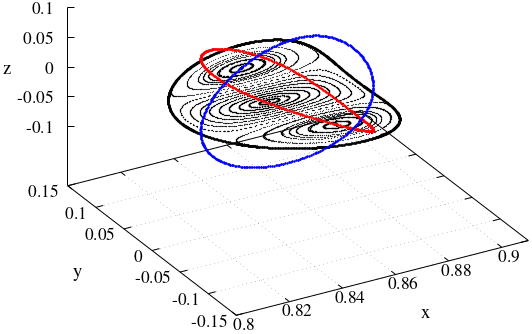}~\includegraphics[width=0.5\columnwidth]{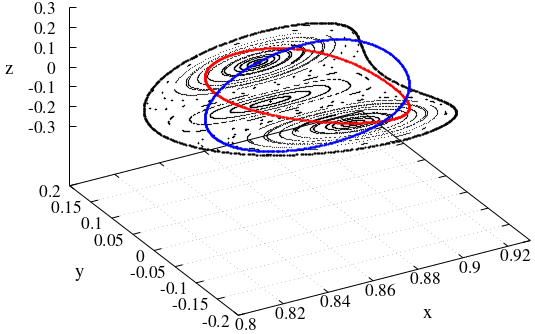}
  \caption{Representation in the space of the Cartesian variables
    $x,y,z$ of the Poincar\'e sections and the projection of the
    (southern and northern) families of the corresponding halo tori
    computed for $f=0$ and for $\kappa=0.025,0.05$ (left and right
    panels respectively).}
\end{figure}

Therefore the halo orbits obtained for large order approximations of
the ERTBP, as it happens with the Lyapunov orbits, are quasi-periodic
orbits belonging to 2-dimensional tori. In figure \ref{twohalos} we
represent in the space of the Cartesian variables $x,y,z$ the
projections of both the Poincar\'e section and the (southern and
northern) families of the corresponding halo tori computed for $f=0$
and for $\kappa=0.025,0.05$ (left and right panels respectively).  As
expected, the halo section of the halo tori for $f=0$ cross the
Poincar\'e section in the corresponding fixed point.

\subsection{Validation of the halo orbits and the halo manifold tubes}

\begin{figure}
  \includegraphics[width=0.8\columnwidth]{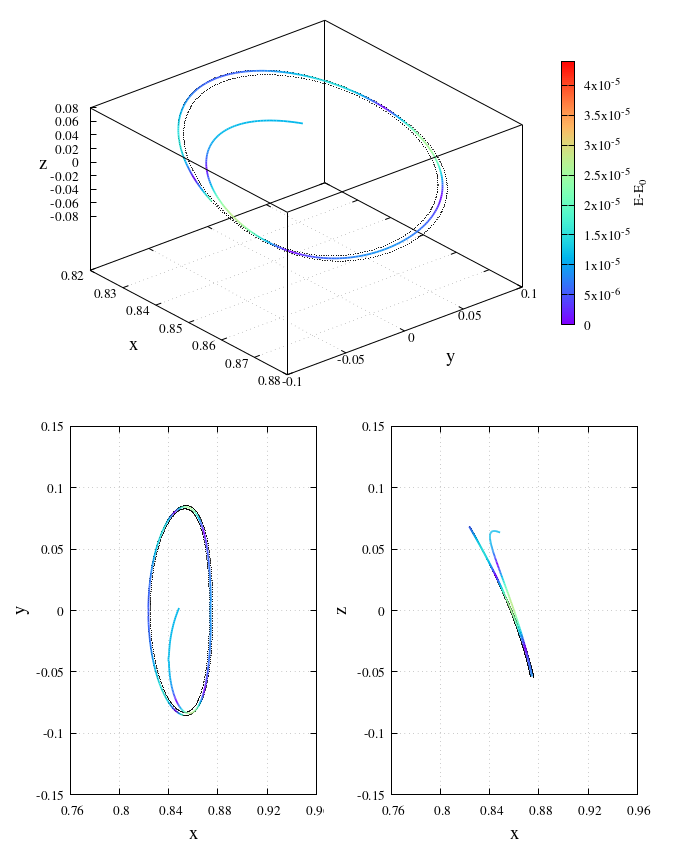}
  \caption{\small A solution of the ERTBP computed by numerically
    integrating the Hamilton equations of (\ref{eq:ori3bp}) with an
    initial condition on the northern halo torus, for
    $\kappa=0.0025$. The black points belong to sections of the torus
    computed for $f=0$ (the external one) and $f=\pi$ (the internal
    one). The numerically computed orbit (colored curve) moves close
    the $f$-section of the Halo torus before departing exponentially
    from it. The upper panel represents the orbit in the $xyz$ space,
    the lower panels the projections on the $xy$ (left) and $xz$
    (right) planes. The color on the orbits represent the variation of
    the local energy, providing an estimate of the neglected remainder
    of the Floquet-Birkhoff normal form along the
    solution.}\label{fig:halok0025}
  \end{figure}

\begin{figure}
   \includegraphics[width=0.5\columnwidth]{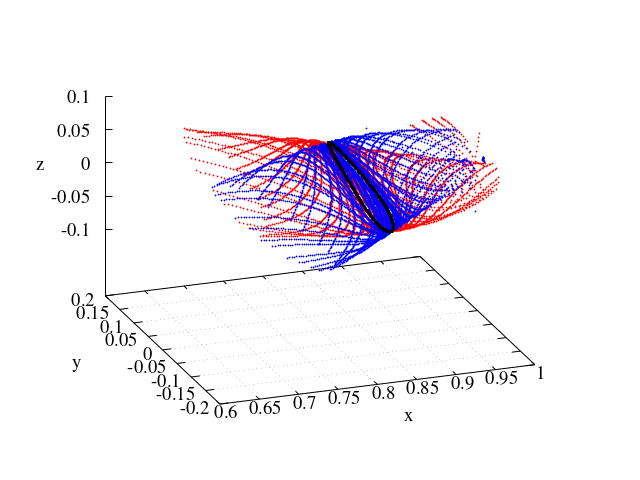}~\includegraphics[width=0.5\columnwidth]{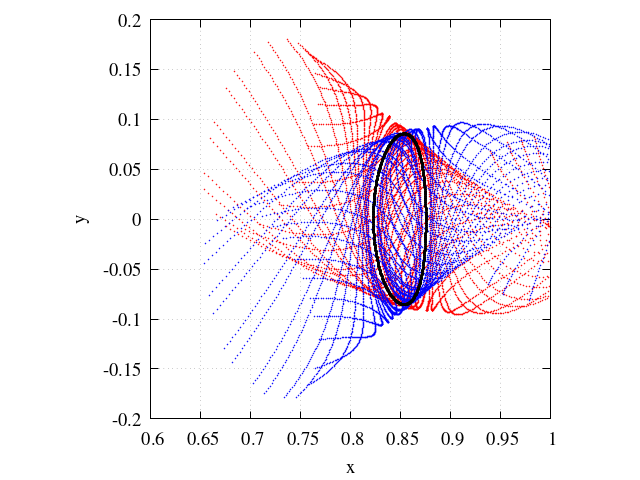}
   \includegraphics[width=0.5\columnwidth]{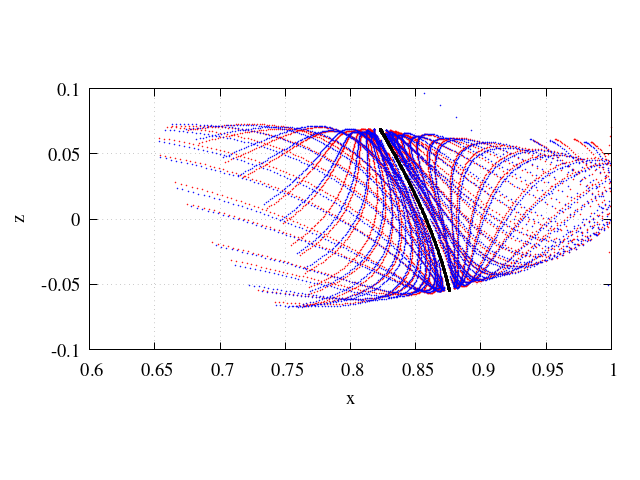}~\includegraphics[width=0.5\columnwidth]{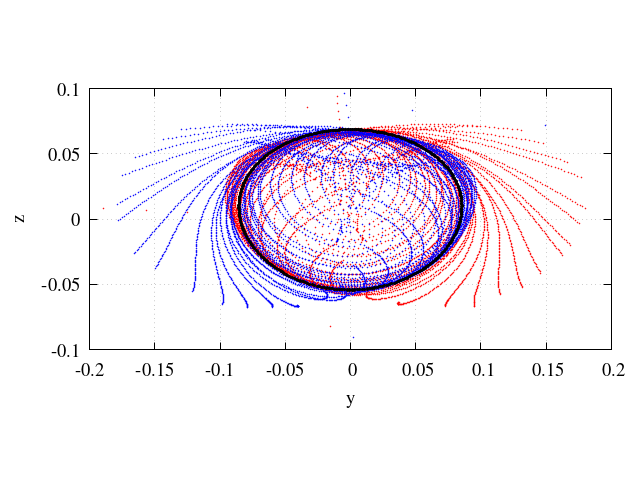}
\caption{\small A sample of numerically computed orbits in the stable
  (blue) and unstable (red) halo tubes, whose initial conditions have
  been obtained using the Floquet-Birkhoff normal form, for $\kappa
  =0.025$. The black dots are in the halo torus.}
\label{fig:tubes}
\end{figure}

In order to test that the orbits generated from the normal form
computations are a good representation of the dynamics of the full
ERTBP, and to determine the effects of eventual small errors, we
perform the following validation test. Along with the section of the
halo tori for $f=0$, we first compute the sections of the torus for
other values of $f$, namely $f=\pi$, $f=\pi/2$, using the Hamiltonian
flow of $K_{CM}$. When transformed to Cartesian variables $x,y,z$,
these sections provide a segmented depiction of the halo torus in the
3 dimensional space.  Then, we numerically integrate the initial
conditions of the halo orbits (transformed to Cartesian variables)
using a numerical integrator of the full ERTBP represented by
Hamiltonian (\ref{eq:ori3bp}). Since the halo torus is hyperbolic in
the normal form dynamics, we expect that the small errors introduced
by neglecting the remainder $R_{N+1}$ are responsible of an hyperbolic
drift of the numerically computed orbit from the analytically computed
halo torus.  As usual with hyperbolic dynamics, even if for small
values of the norm of $R_{N+1}$ the errors on the initial conditions
are small, this small error grows exponentially in time. As it happens
for the computation of the Lyapunov orbits of the CRTBP, the
hyperbolic components is so strong that typically the numerically
integrated orbits depart from the analytically computed ones within
few periods. We check for how long the evolution of the numerically
computed halo orbits remains close to the corresponding analytically
computed halo torus for a value of the local energy $\kappa =0.025$.
In Figure \ref{fig:halok0025} we represent a numerically
integrated orbit which remains close to the torus for a full
circulation before departing exponentially from it. As expected, the
larger is the local energy energy, the larger is the amplitude of the
corresponding halo torus and the shorter is the time required for the
orbit to depart from it. We also represent with a color scale the
variation of the value of the local energy $\kappa$ during the
numerical integration, which represents the best
  estimator of the error. In fact, the variation of the local energy
  remains small also when the orbit departs form the torus.  The
exponential instability of individual orbits provides
  an opportunity to construct orbits of the full ERTBP which arrive
close to (or depart from) the halo torus or that transit close to it,
as it has been done in correlation with the Lyapunov tori of the ERTBP
previously studied in \cite{PG21}. A sample of numerically computed
orbits in the stable and unstable halo tubes, whose initial conditions
have been obtained using the Floquet-Birkhoff normal form, is
represented in Figure \ref{fig:tubes}.

\subsection{The remainder $R_{N+1}$}\label{sec:remainder}

Another validation test is performed with a direct computation of the
norm of the remainder $R_{N+1}$ of the Floquet-Birkhoff normal form
(\ref{eq:BTnf}) along the halo orbit, as well as in a
neighbourhood. In fact, since the halo tori are characterized by large
librations from the corresponding Lagrangian solution, it is important
to check that the remainder $R_{N+1}$ (which is neglected in the
definition of the torus) is indeed small in a neighbourhood of the
torus.  Previously we have checked with an indirect method the norm of
the remainder by computing the variation of the local energy along a
numerically integrated solution of the ER3BP. Now we represent a
direct computation of the norm of the remainder computed along
different halo tori and different normalizations orders.

We first define the maximum of the norm of the remainder on
a set of points ${\cal S}$ as
\begin{equation} 
  \norm{R^{(J)}} :=
  \mathrm{Max}_{(\hat{\mathbf{q}},\hat{\mathbf{p}},f) \in {\cal S}}
  \sum_{j=J+1}^{10}
  \norm{\hat{H}_j^{(J)}(\hat{\mathbf{q}},\hat{\mathbf{p}},f)}~,
  \label{eq:normrem}
\end{equation}
for all the normalization orders $J=2,\ldots,N$, and then we compute
it on a set ${\cal S}$ of points sampling the Halo orbits.  The
results are summarized in Table~\ref{tab:rem}, and show the orders of
magnitude of improvement in the error of our best Floquet-Birkhoff
normal form (of order $J=8$) with respect to the classical Floquet
approximation where no Birkhoff transformations are implemented
(corresponding to order $J=2$).

\begin{table}[h]
\centering
\begin{tabular}{|c|c|c|c|c|c|}
  \hline \hline
  \multicolumn{6}{|c|}{$\mu = 0.012300$ $e = 0.0549006$ (Earth-Moon system)}\\
  \multicolumn{6}{|c|}{Center manifold Halos}\\
  \hline
  \multicolumn{2}{|c|}{Halo for $\kappa=0.025$} &
  \multicolumn{2}{|c|}{Halo for $\kappa=0.030$} &
  \multicolumn{2}{|c|}{Halo for $\kappa=0.050$} \\ 
  \hline
  $J$ & $|R^{(J)}|$ & $J$ & $|R^{(j)}|$ & $J$ & $|R^{(j)}|$ \\
  \hline
  $2$ & $1.45043\snot[-2]$ & $2$ & $2.39118\snot[-2]$ & $2$ & $1.12098\snot[-1]$ \\
  $3$ & $1.42105\snot[-3]$ & $3$ & $2.39449\snot[-3]$ & $3$ & $1.08179\snot[-2]$ \\
  $4$ & $5.40381\snot[-4]$ & $4$ & $9.88391\snot[-4]$ & $4$ & $5.55895\snot[-3]$ \\
  $5$ & $2.58946\snot[-4]$ & $5$ & $5.10725\snot[-4]$ & $5$ & $3.48433\snot[-3]$ \\
  $6$ & $1.09358\snot[-4]$ & $6$ & $2.33717\snot[-4]$ & $6$ & $1.96300\snot[-3]$ \\
  $7$ & $4.41467\snot[-5]$ & $7$ & $1.01352\snot[-4]$ & $7$ & $1.02467\snot[-3]$ \\
  $8$ & $1.47660\snot[-5]$ & $8$ & $3.60189\snot[-5]$ & $8$ & $4.25978\snot[-4]$ \\
  \hline
  \hline
\end{tabular}
\caption{Estimation of the norm of the
  remainder~\eqref{eq:normrem} for the three halo orbits of previous section,
  for energies $\kappa=0.025$, $\kappa=0.030$, $\kappa=0.050$.}
\label{tab:rem}
\end{table}

\begin{figure}
\includegraphics[width=1\columnwidth]{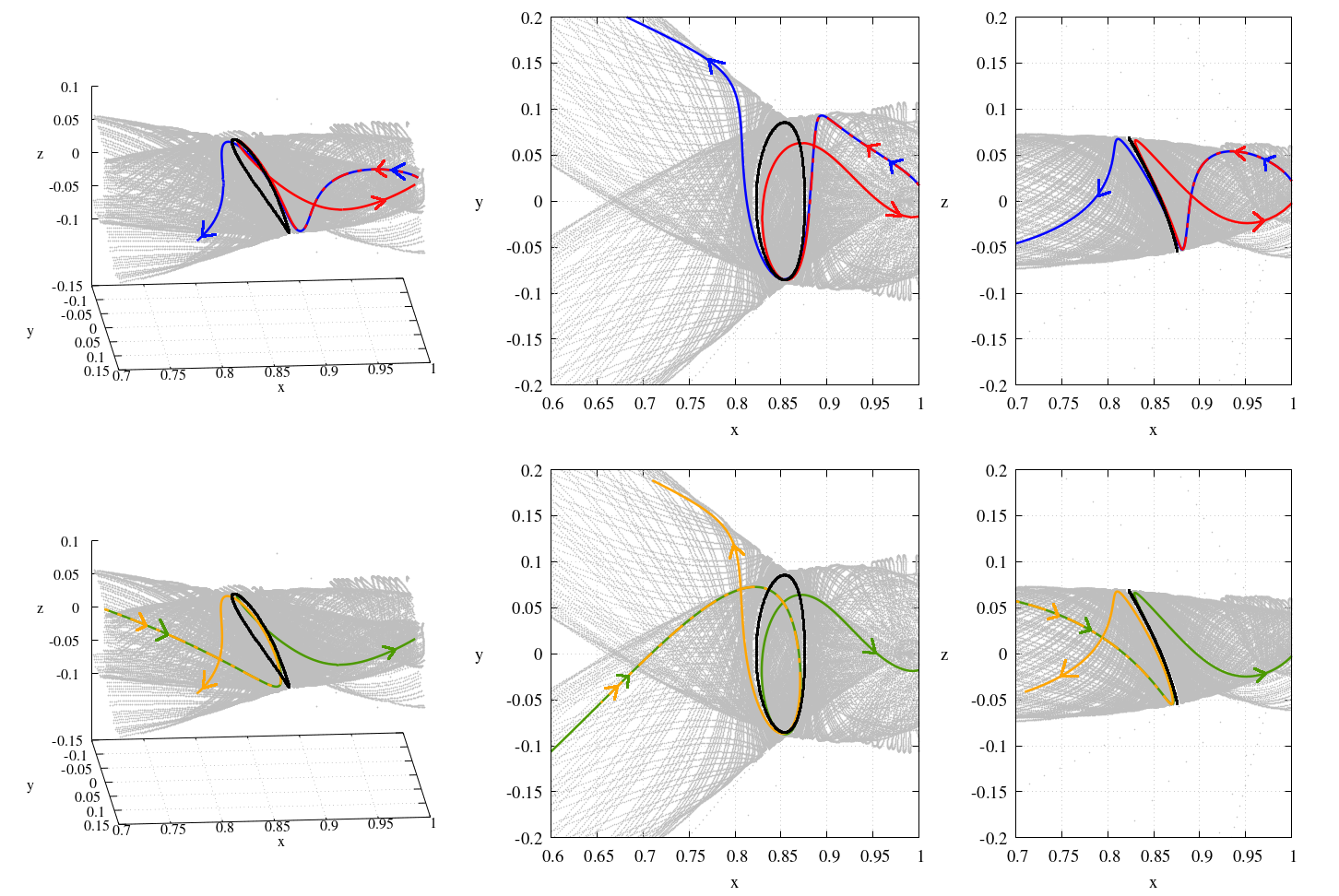}       
\caption{\small Numerically computed orbits in the stable and unstable
  halo tubes (in gray), together with orbits which transit from one
  side of the halo torus to the other, or bounce back. The initial
  conditions of all these orbits have been obtained using the
  Floquet-Birkhoff normal form, for $\kappa =0.025$, the black dots
  are in the halo torus.}
\label{fig:tubestransits}
\end{figure}

Let us remark that the an effect of the error introduced by
truncating the remainder is that for the orbits with initial
conditions which are on the halo orbits evolve
as the orbits which are in a small neighbourhood of the stable and unstable
manifold tubes. According the the position of the 
initial values of the hyperbolic variables with respect to
the values of the stable and unstable tubes provides orbits
which have different transit properties at the halo orbits. 
In figure \ref{fig:tubestransits} we represent orbits
which are very close to the stable and unstable tubes, but after they
approach the halo torus, depending on their position relative to the
tubes they transit from one side to the other of the halo torus,
or instead they bounce back. We therefore find the same kind of
transits behaviour that has been previously found for the
planar Lyapunov orbits. 

\section*{Acknowledgments}

The authors acknowledge the project MIUR-PRIN 20178CJA2B "New frontiers
of Celestial Mechanics: theory and applications".




\end{document}